# Gain characteristics of a 100 μm thick GEM in Krypton-$CO_2$ mixtures

R. C. Roque[a], H. Natal da Luz[b], L. F. N. D. Carramate[c], C. D. R. Azevedo[c], J. A. Mir[a], F. D. Amaro[a*]

[a] *LIBPhys  Department of Physics, University of Coimbra*
  *3004-516 Coimbra, Portugal*

[b] *Instituto de Física, Universidade de São Paulo*
  *Brasil*

[c] *I3N - Physics Department, University of Aveiro*
  *3810-193 Aveiro, Portugal*
   *E-mail*: `famaro@gian.fis.uc.pt`

ABSTRACT: The operation of a Micropattern Gaseous Detector (MPGD) comprising a cascade of two non-standard Gas Electron Multipliers (GEM) made from a 100 μm kapton foil was evaluated in pure krypton and in krypton-$CO_2$ mixtures. The final avalanche charge was collected in a 2D strip readout, providing an active area of $10\times10$ $cm^2$ for the entire detection system. For each mixture, the effective gain and energy resolution were measured as a function of the electric field in the drift, transfer and induction regions. Using X-rays from a $^{55}$Fe radioactive source (5.9 keV), gains close to $10^4$ and energy resolution of 22% (FWHM) were achieved in pure krypton under stable conditions, showing the potential of the experimental setup for the following imaging studies.

# Contents

## 1. Introduction

The Gas Electron Multiplier (GEM) was invented at CERN and is manufactured using photolithographic methods and etching techniques [1]. A typical GEM plate is made from a 50 µm thick, copper clad, kapton foil perforated with a hexagonal pattern of bi-conical holes [2]. When immersed in an adequate gaseous volume and under the influence of suitable electric fields across the electrodes, it is able to multiply the charge that goes through its holes [1]. Higher amplifications are possible using a cascade of GEM plates that act as phased charge multipliers. Also, it has the ability to preserve the 2D spatial information of the initial ionization achieving resolutions down to 100 µm and making it adequate for imaging applications [3].

However, typical GEM foils are still sensitive to sparking and can be permanently damaged after a significant discharge. Thick-GEM (THGEM) structures, usually made of 0.4 mm thick G10 plates with micro drilled holes, were developed in order to provide a more robust charge amplifying stage to operate under critical conditions. The gains measured with this alternative method are, for the same applied voltage and gas mixture, lower than the ones achieved with typical GEM plates but comparable performances can be obtained [4].

The GEM structures used in this work, GEM-100, were fabricated at CERN using the same wet chemical etch process used for the production of the standard (50 µm thick) GEM [5]. They were made from a 100 µm thick kapton foil and have a hole diameter and pitch of 100 µm and 200 µm, respectively. Previous works showed that gain charges of $3\times10^3$ and $10^4$ were achieved in single and double stage configurations, respectively, using an $Ar-CO_2$ (70:30) mixture. These results are similar to the ones recorded using standard GEMs in the corresponding configurations, but presented higher robustness to sparking [5].

In a follow up study [6] an imaging detector with an active area of $10\times10$ cm$^2$ and comprising the same GEM plates was developed. Immersing the detector in an $Ar-CO_2$ mixture, an energy resolution of 21% and charge gains above $10^4$ were recorded using a $^{55}$Fe radioactive source. Gain and energy resolution 10% and 4.9% higher than the average values were, respectively, recorded at the edges of the detector. Position resolution was also evaluated from acquired images and a minimum value of 1.7 mm was achieved when irradiating the detector with X-rays in the range 10 keV to 12 keV.

The goal of this work is to investigate the performance of the detector used in the previous studies operating in pure krypton and Kr-$CO_2$ mixtures. For energies in the range 17 to 25 keV, krypton presents the best value of position resolution amongst the noble gases, 18 to 50 µm [7][8], being a good candidate for imaging detectors. In this work we measure the charge gain and energy resolution of a cascade of two GEM-100 operating in pure krypton and in two Kr-$CO_2$ mixtures (90:10 and 80:20).

## 2. Experimental Setup

Figure 1 depicts the main characteristics of the experimental setup used in this study; while figure 1a represents the position of the different elements in the detector and its dimensions, figure 1b is a schematic of the 2D strip readout with resistive lines connecting the strips responsible for charge collection. The GEM-100 foils were made from a 100 µm thick kapton foil which is twice the thickness of a standard GEM. A cascade setup was chosen in order to provide two multiplication stages to increase the signal to noise ratio.

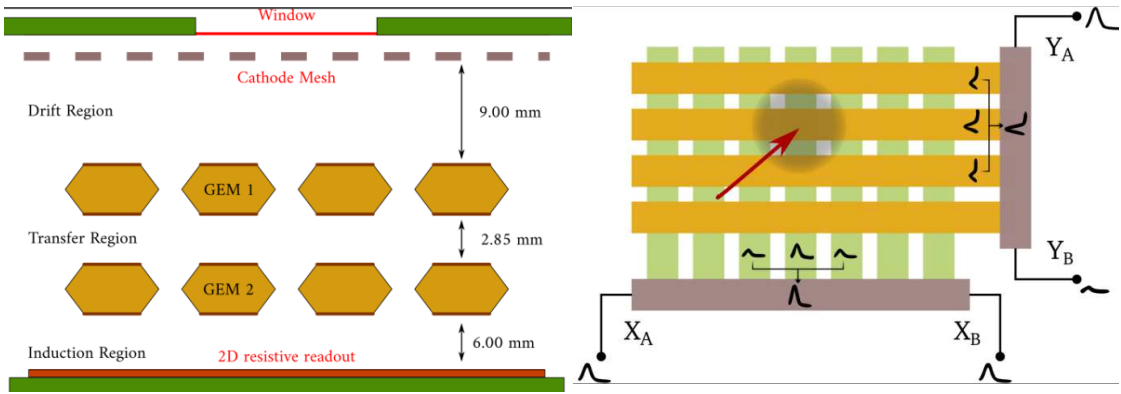

Figure 1: (Left) Schematics of the detector used; a cascade of two 100 µm GEM foils provides two multiplication stages while a metallic mesh limits the drift region. (Right) 2D readout; two perpendicular sets of strips allow the recording of position and event energy. Each set of strips is connected by a resistive line. The amplitude of the signals recorded in each end of the resistive line allows to do the 1D position reconstruction (x or y). For the results presented in this work the 4 channels of the 2D readout were interconnected and no information on the interaction position was recorded.

The GEM-100, the cathode mesh used to establish the drift field and the 2D readout have an active area of 10×10 $cm^2$. The cathode mesh is made of 80 µm diameter stainless steel wires with 900 µm spacing. While the 2D readout was grounded, the GEM-100 electrodes and the cathode mesh were independently biased with negative voltages, using CAEN V6521 HN power supplies with current limitation of 100 nA. Each channel was additionally filtered using a custom made low pass filter with RC constant of 20 ms.

Two orthogonal sets of conductive strips, each connected by a resistive line, compose the 2D readout, allowing the recording of the position where each interaction takes place as well as the deposited energy [9]. For the gain and energy resolution studies presented in this work, the four output channels of the 2D readout ($X_A$, $X_B$, $Y_A$ and $Y_B$ in figure 1b) were short-circuited and no information of the position was recorded. A charge sensitive preamplifier (Canberra 2004) was directly connected to the 2D readout output. It fed a linear amplifier Tennelec (TC-243) with 4 µs peaking time whose output was digitalized by an analog to digital converter (Nucleus PCA). The electronic chain sensitivity was calibrated by injection of a known charge into the preamplifier input.

A $^{55}$Fe source was used to perform all the measurements presented in this work. The source was collimated to a diameter of 2 mm and orthogonally positioned relatively to the detector window which was made of a thin (5 µm) film of aluminized Mylar. The Mn K$\alpha$ and K$\beta$ lines (5.89 keV and 6.49 keV, respectively) were both present at the measurements for the electric field optimization, the latter with intensity 9 times smaller than the former.

Pure krypton and two gas mixtures, having krypton as main component, Kr-$CO_2$ (90:10) and Kr-$CO_2$ (80:20) were studied. For all measurements a constant gas flow of 2 l/h, corresponding to an exchange rate of one gas volume per hour, was held. The pure krypton and $CO_2$ were mixed using mass flow controllers EL-FLOW Prestige (Bronkhorst). For each mixture, the effective gain and energy resolution were studied as a function of voltage differences between the various electrodes of the system.

## 3. Results

The performance of the double GEM-100 cascade was evaluated in the above mentioned gas mixtures having krypton as main component. Effective charge gain and energy resolution (relative to the first data taken for each measurement) were calculated as a function of the electric field established in the drift, transfer and induction regions, as depicted in figure 1a. The results are described in the following subsections.

### 3.1 Drift Field Optimization

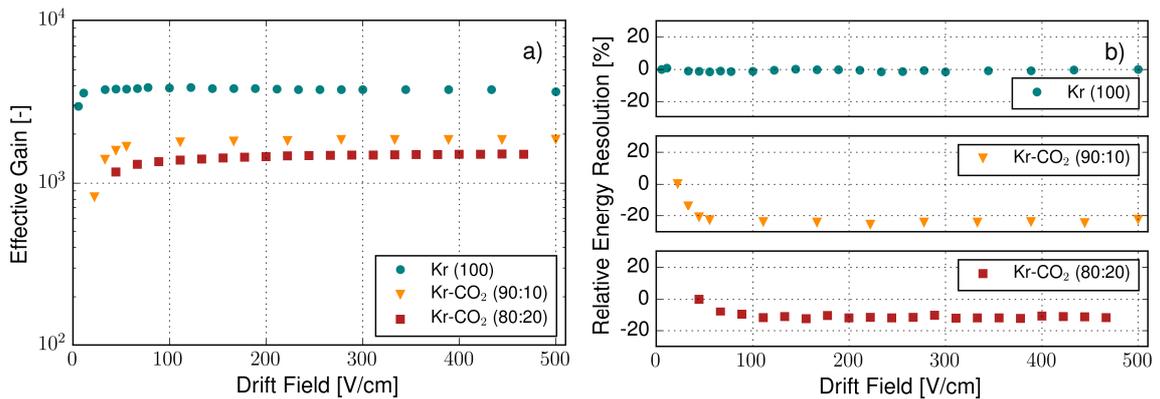

Figure 2: Effective charge gain (a) and relative energy resolution (b) as a function of the electric field in the drift region, for the gas mixtures studied in this work. The energy resolution is relative to the first data point taken.

For the drift field optimization, the voltage between the cathode mesh and the top of the first GEM was modified while the other electrodes were kept at constant potentials. The voltage between the top and bottom electrodes of the GEM-100, $\Delta V_{GEM}$, was optimized for each gas mixture, taking into account the discharge limits of the GEM foils. It was 550 V across both GEM's in the 90:10 measurements and 600 V for the 80:20 mixture. For pure krypton these values were set at $\Delta V_{GEM1}$ = 530 V and $\Delta V_{GEM2}$ = 560 V.

Figure 2 (a) shows that pure krypton promotes a faster rise to plateau where effective gain and energy resolution remain constant and that lower fields are required to achieve full primary electron collection. The inclusion of $CO_2$ in the mixture increases the electric field required to assure full collection of the primary electric charges. The plateau is reached at 50 V×cm$^{-1}$ in pure krypton, 150 V×cm$^{-1}$ in Kr-$CO_2$ (90:10) and 300 V×cm$^{-1}$ in Kr-$CO_2$ (80:20).

### 3.2 Transfer Field Optimization

During this part of the study, the electric field between the bottom of the first GEM and the top of the second one, $E_{transf}$, was changed while the remaining electric fields across the detector were kept constant. The transfer field plays a decisive role in the extraction of the charge from the holes of the first GEM and their focusing into the holes of the second GEM. The requirements for these two actions are opposite: while a large transfer field is required for adequate extraction of the electrons from the holes of the first GEM of the cascade, increasing the transfer field decreases the focusing of the electrons into the holes of the second GEM, causing the electrons to be trapped at the top electrode of the second GEM, and thus reducing the overall gain of the cascade. Figure 3 shows the effective gains and energy resolution for each of the mixtures studied, as a function of the transfer field.

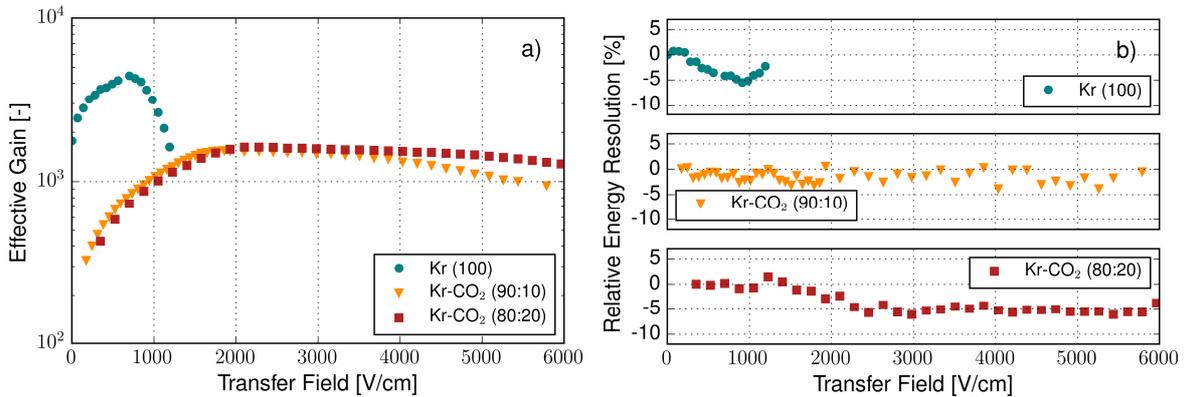

Figure 3: Effective gain (a) and relative energy resolution (b) as a function of the transfer field for pure krypton, krypton-$CO_2$ (90:10) and krypton-$CO_2$ (80:20). The energy resolution is relative to the first data point taken.

We have observed that the operation in pure krypton is more sensitive to small changes in the transfer field since no plateau where the effective gain and energy resolution remain constant is reached. Figure 3 shows that the gains in pure krypton reach values above $4\times10^3$ for $E_{transf}$ = 560 V×cm$^{-1}$ and then dropping rapidly to lower values. The krypton-$CO_2$ mixtures present a different behavior, with a much less severe drop in the gain with increasing $E_{transf}$. They have a similar profile to each other with a maximum gain about $1.6 \times 10^3$ and a plateau of 1 kV×cm$^{-1}$ where the effective gain does not change significantly, since the effect of loss of collection efficiency in the second GEM is compensated by the increase of the extraction efficiency from the first one.

### 3.3 Induction Field Optimization

During the optimization of the electric field in the induction region all the other electric fields were kept constant. This region is responsible for extracting the electrons from the holes of GEM2 and guide them to the readout board. In order to collect the maximum possible charge a high electric field is required. The results are shown in figure 4.

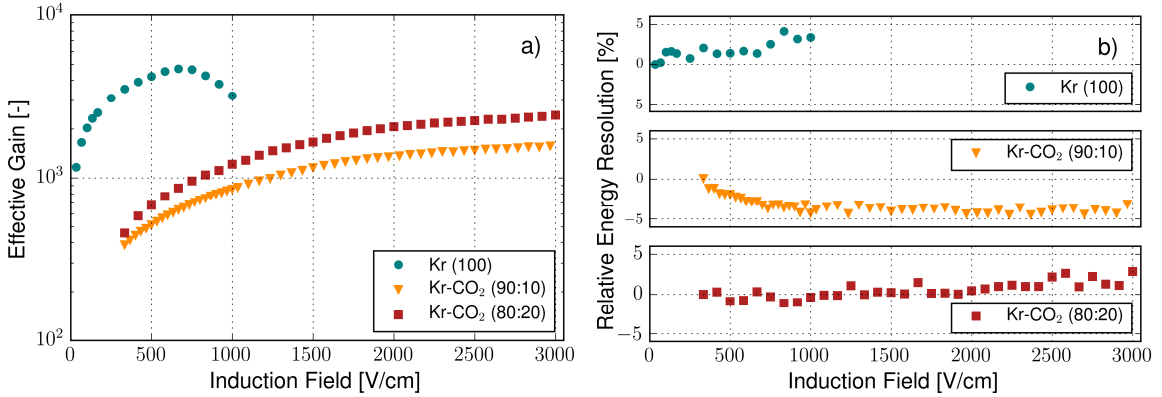

Figure 4: Effective gain (a) and relative energy resolution (b) as a function of the induction field for pure krypton, Kr-CO$_2$ (90:10) and Kr-CO$_2$ (80:20). The energy resolution is relative to the first data point taken.

For the Kr-CO$_2$ mixtures, extraction from the holes of the second GEM-100 increases with increasing electric field, as expected. For the 90:10 mixture, an effective gain around $1.5 \times 10^3$ was measured at an induction field of 3.0 kV×cm$^{-1}$. For the 80:20 mixture, gains of $2.4 \times 10^3$ where achieved. Higher extraction fields could not be reached due to the increasing voltages that must be applied to the electrodes of the detector in order to keep all the remaining electric fields constant.

However, for pure krypton, a decrease of the effective gain occurs when electric fields above 700 V×cm$^{-1}$ are applied, providing a maximum near $4.7 \times 10^3$. This behavior is not easily explained. A repetition of the measurements was performed with similar results.

### 3.4 GEM Optimization

Following the optimization of the electric fields in the drift, transfer and extraction regions, we have studied the gain and energy resolution of the double GEM-100 cascade as a function of the voltage difference across the GEM foils. For each gas mixture the GEM-100 foils were biased with the same voltage difference, $\Delta V_{GEM}$, and the electric fields in the drift, transfer and extraction regions were selected taking into account the information (effective gain and energy resolution) presented in the previous sections. The electric fields selected for the drift/transfer/induction regions were (in V/cm): pure krypton (200/650/580), Kr-CO$_2$ (90:10) (333/1930/2500), Kr-CO$_2$ (80:20) (310/2800/1500). $\Delta V_{GEM}$ was selected taking into account the onset of discharges. The $^{55}$Fe source used in the previous measurements was further collimated (1 mm diameter) and a thin Cr foil was used to filter the K$\beta$ line.

Figure 5a) shows that the highest charge gain ($1 \times 10^4$) was achieved for the 90:10 mixture when both GEM electrodes where biased with 575 V. For pure krypton and for the 80:20 mixture a maximum gain of $7\times10^3$ was reached just before discharges took over. It is also visible that the higher the concentration of $CO_2$, the higher the voltages required across the GEM plates to achieve the same value of charge gain.

The increase of $CO_2$ in the mixture also implies a degradation in energy resolution; for pure krypton energy resolution lies around 22% as for the 90:10 and 80:20 mixtures values near 25% and 27% were recorded, respectively. Figure 6 presents a histogram of the normalized pulse height amplitudes recorded for pure krypton and the mixture 80:20 mixture. The broadening of the spectrum with increasing $CO_2$ concentration is clearly visible.

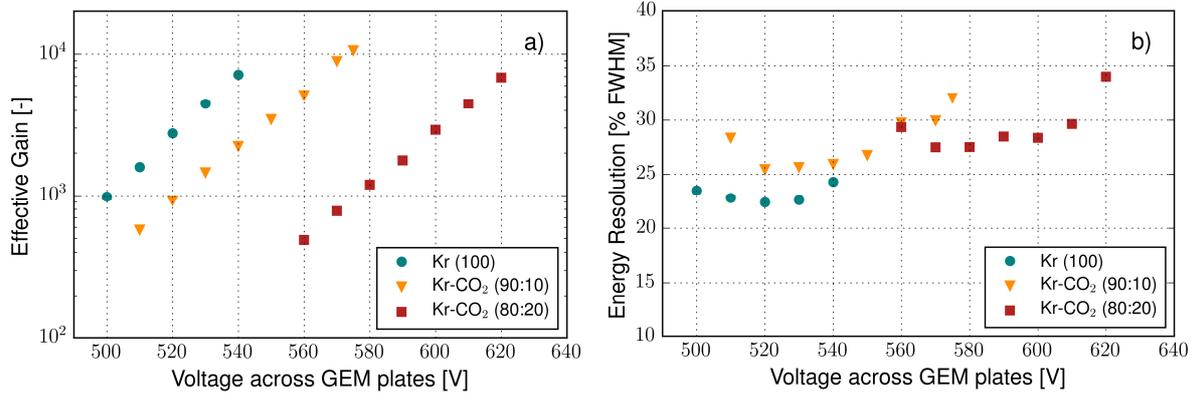

Figure 5: Charge multiplication, (a) and energy resolution, b) for the GEM foils operated in different Kr-$CO_2$ mixtures. The electric fields in the drift, transfer and induction regions were for each mixture, respectively (V×cm$^{-1}$): pure krypton (200/650/580), Kr-$CO_2$ (90:10) (333/1930/2500), Kr-$CO_2$ (80:20) (310/2800/1500).

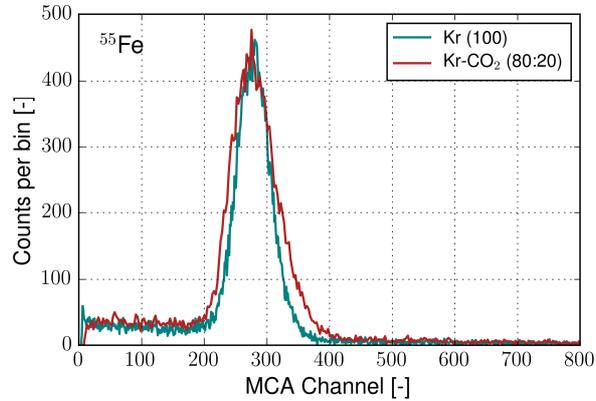

Figure 6: Pulse height distribution recorded for pure krypton and the Kr-$CO_2$ (80:20) mixture. Energy resolutions measured in pure krypton and Kr-$CO_2$ (80:20) were, respectively, 22 and 27% (FWHM, 5.9 keV).

## 4. Conclusions

The charge gain measured in all krypton based mixtures was above $6\times10^3$, reaching a maximum of $1\times10^4$ in the Kr-$CO_2$ (90:10) mixture. These values are compatible with the Ar-$CO_2$ results [6]. For pure krypton, Kr-$CO_2$ (90:10) and Kr-$CO_2$ (80:20), the minimum energy resolutions recorded were 22%, 25% and 27%, respectively. These results prove, once more, the capacity of a 100 µm thick Gas Electron Multiplier to reproduce the standard GEM operation but with higher tolerance to critical conditions such as sparking.

Another relevant observation is the gain behaviour of the double GEM cascade operating in pure krypton as a function of the electric field in the transfer and induction region. Repeated experimental results showed consistently that electron extraction from the last GEM decreased sharply for electric fields above 700 V×cm$^{-1}$; this uncommon result calls for further investigation.

Future developments of this work comprise image acquisition using the evaluated gaseous mixtures and a continuous x-ray generator instead of the mono-energetic $^{55}$Fe source. The position reconstruction will be made using the resistive charge division method, as the current experimental system is already prepared to perform such studies.

**Acknowledgments**

L. F. N. D. Carramate, C. D. R. Azevedo and F. D. Amaro acknowledges FCT grants SFRH/BPD/120571/2016, SFRH/BPD/79163/2011 and SFRH/BPD/74775/2010, respectively. H. Natal da Luz acknowledges grant 2016/05282-2, São Paulo Research Foundation (FAPESP). Work/research carried out within the R&DT project CERN/FP/123614/2011. This project was developed under the scope of a QREN initiative, UE/FEDER financing, through the COMPETE programme (Programa Operacional Factores de Competitividade).

**References**

[1] F. Sauli, *Gem: A new concept for electron amplification in gas detectors*, Nuclear Instruments and Methods A 386 (1997) 531.

[2] A. F. Buzulutskov, *Radiation detectors based on gas electron multipliers (review),* Instruments and Experimental Techniques 50 (2007) 287.

[3] W. Dabrowski, et al., *Application of GEM-based detectors in full-field XRF imaging*, 2016 *JINST* 11 C12025.

[4] R. Chechik, A. Breskin, C. Shalem, *Thick gem-like multipliers - a simple solution for large area uv-rich detectors*, Nuclear Instruments and Methods A 553 (2005) 35.

[5] J. A. Mir, et al., *Gain characteristics of a 100 µm thick Gas Electron Multiplier (GEM),* 2015 *JINST* 10 C12006.

[6] F. D. Amaro, et al., *A robust large area x-ray imaging system based on 100 µm thick Gas Electron Multiplier*, 2015 *JINST* 10 C12005.

[7] J. Fischer, V. Radeka, G.C. Smith, *Position detection of 17-25 keV x-rays in krypton and xenon with a resolution of 18-50 µm (FWHM)*, IEEE Transactions on Nuclear Science 33 (1986) 257.

[8] C. D. R. Azevedo, et al., *Position resolution limits in pure noble gaseous detectors for X-ray energies from 1 to 60 keV,* Physics Letters B 741(2015) 272.

[9] H. Natal da Luz, et al., *X-ray imaging with GEMs using 100 µm thick foils*, 2014 *JINST* 9 C06007.